# Prototype of the SST-1M Telescope Structure for the Cherenkov Telescope Array

**J. Niemiec**[*,b], **W. Bilnik**[a], **J. Błocki**[b], **L. Bogacz**[c], **J. Borkowski**[f], **T .Bulik**[d], **F. Cadoux**[e], **A. Christov**[e], **M. Curyło**[b], **D. della Volpe**[e], **M. Dyrda**[b], **Y. Favre**[e], **A. Frankowski**[f], **Ł. Grudnik**[b], **M. Grudzińska**[d], **M. Heller**[e], **B. Idźkowski**[g], **M. Jamrozy**[g], **M. Janiak**[f], **J. Kasperek**[a], **K. Lalik**[a], **E. Lyard**[h], **E. Mach**[b], **D. Mandat**[i], **A. Marszałek**[b], **J. Michałowski**[b], **R. Moderski**[f], **T. Montaruli**[e], **A. Neronov**[h], **M. Ostrowski**[g], **P. Paśko**[j], **M. Pech**[i], **A. Porcelli**[e], **E. Prandini**[h], **P. Rajda**[a], **M. Rameez**[e], **E. jr Schioppa**[e], **P. Schovanek**[i], **K. Seweryn**[j], **K. Skowron**[b], **V. Sliusar**[k], **M. Sowiński**[b], **Ł. Stawarz**[g], **M. Stodulska**[g], **M. Stodulski**[b], **I. Troyano Pujadas**[e], **S. Toscano**[h,l], **R. Walter**[h], **M. Więcek**[a], **A. Zagdański**[g], **K. Ziętara**[g], **P. Żychowski**[b], and **M. Kłaczyński**[a], **R. Kocierz**[a], **P. Kuras**[a], **P. Kurowski**[a], **Ł. Ortyl**[a], **T. Owerko**[a]

[a]*AGH University of Science and Technology, al.Mickiewicza 30, 30-059 Kraków, Poland;*
[b]*Instytut Fizyki Jądrowej im. H. Niewodniczańskiego Polskiej Akademii Nauk, ul. Radzikowskiego 152, 31-342 Kraków, Poland;* [c]*Department of Information Technologies, Jagiellonian University, ul. Reymonta 4, 30-059 Kraków, Poland;* [d]*Astronomical Observatory, University of Warsaw, al. Ujazdowskie 4, 00-478 Warsaw, Poland;* [e]*DPNC - Université de Genéve, 24 Quai Ernest Ansermet, CH-1211 Genéve, Switzerland;* [f]*Nicolaus Copernicus Astronomical Center, Polish Academy of Sciences, ul. Bartycka 18, 00-716 Warsaw, Poland;* [g]*Astronomical Observatory, Jagiellonian University, ul. Orla 171, 30-244 Kraków, Poland;* [h]*ISDC, Observatoire de Genéve, Université de Genéve, 16 Chemin de Ecogia, CH-1290 Versoix, Switzerland;* [i]*Institute of Physics of the Czech Academy of Sciences, 17. listopadu 50, Olomouc & Na Slovance 2, Prague, Czech Republic;* [j]*Centrum Badań Kosmicznych Polskiej Akademii Nauk, 18a Bartycka str., 00-716 Warsaw, Poland;* [k]*Astronomical Observatory, Taras Shevchenko National University of Kyiv, Observatorna str., 3, Kyiv, Ukraine; also with* [l]*Vrije Universiteit Brussels, Pleinlaan 2 1050 Brussels, Belgium*
*E-mail:* Jacek.Niemiec@ifj.edu.pl

**for the SST-1M sub-consortium and the CTA Consortium**[†]

A single-mirror small-size (SST-1M) Davies-Cotton telescope with a dish diameter of 4 m has been built by a consortium of Polish and Swiss institutions as a prototype for one of the proposed small-size telescopes for the southern observatory of the Cherenkov Telescope Array (CTA). The design represents a very simple, reliable, and cheap solution. The mechanical structure prototype with its drive system is now being tested at the Institute of Nuclear Physics PAS in Krakow. Here we present the design of the prototype and results of the performance tests of the structure and the drive and control system.

*The 34th International Cosmic Ray Conference,*
*30 July- 6 August, 2015*
*The Hague, The Netherlands*

---

[*]Speaker.
[†]Full consortium author list at http://cta-observatory.org

## 1. Introduction

Small-size telescopes (SSTs) are conceived to cover the highest energy range of the Cherenkov Telescope Array (CTA) observatory between a few TeV and 300 TeV. A consortium of Polish and Swiss institutions has recently built a prototype mechanical structure of the single-mirror (SST-1M) telescope that utilizes a standard and proven Davies-Cotton design. It will be equipped with an innovative fully digital and lightweight camera based on Silicon Photomultipliers (SiPM). The SST-1M has a focal length of 5.6 m, a pixel angular opening of 0.24°, a field of view of 9°, and a reflective dish with diameter of 4 m. The dish is composed of 18 hexagonal mirror facets with size of 78 cm flat-to-flat and offers a total effective collection area of 9.42 m$^2$. The latter is reduced to 6.47 m$^2$ after correcting for the shadowing due to mast and camera chassis and taking into account the reflectance of the mirrors.

The SST-1M is particularly attractive since it has a simple, compact, and lightweight mechanical structure with the total weight of about 8.6 tons, is cheap to produce and easy to construct, install, and maintain. Yet the design is stiff and solid, suitable for sites well above 2 km in altitude and able to resist the earthquake conditions of the Chilean site proposed for the southern array of CTA. The synergy with the Medium-Size Telescope (MST) for CTA, achieved in terms of the drive system components and the control software, provides the uniformity across CTA telescope array sub-components that should markedly simplify the maintenance of CTA.

The prototype SST-1M telescope structure was installed at INP PAS in Kraków in November 2013. This work presents the tests of the telescope structure and the drive and control system that validate the SST-1M as a viable solution for CTA.

## 2. Prototype of the SST-1M mechanical structure

### 2.1 Frame

The mechanical prototype of SST-1M was built mainly to test tendering process steps, fabrication processes, manufacturing tolerances, assembly, mounting and installation, commissioning, and also mirror mounting procedure. It was constructed according to technical documentation based on the structure design described in detail in [1]. The frame manufacturing and installation was done by the Ponar Sp. z o.o. company from Żywiec, Poland. The control cabinet was integrated at INP PAS. The material used for the structure is mostly steel and all steel profiles and tubes are off-the-shelf products from industry. Figure 1 shows the prototype and the control cabinet. Note that dummy mirrors and mirror interfaces with appropriate weights and sizes have been used. In addition, steel plates attached to the camera-holding frame served as a substitute of the camera weight. This was needed to properly balance the telescope before tests could be performed. 15 out of 18 mirror facets were installed on the prototype in June, 2015.

### 2.2 Drive and drive control and safety systems

The telescope positioning and tracking system is realized with two independent drive axes: the azimuth axis and the elevation axis. The movement of the telescope around each of these axes is realized with a system of one slew drive and one roller bearing [1]. The slew drive is a compact system that combines a worm gear with a motor and also a roller bearing, thus enabling transmission of both radial and axial forces. It has a fully enclosed and self-supporting housing. The SST-1M design adopts an IMO slew drive of the WD-H type, the same type as for medium-size telescope (MST). Each slew drive contains a twin worm gear with two servo-motors. Such a solution helps increasing the torque capacity, eliminates backlash ensuring a higher pointing accuracy, increases the reliability of the system, and provides higher flexibility that can be used if high acceleration is needed.

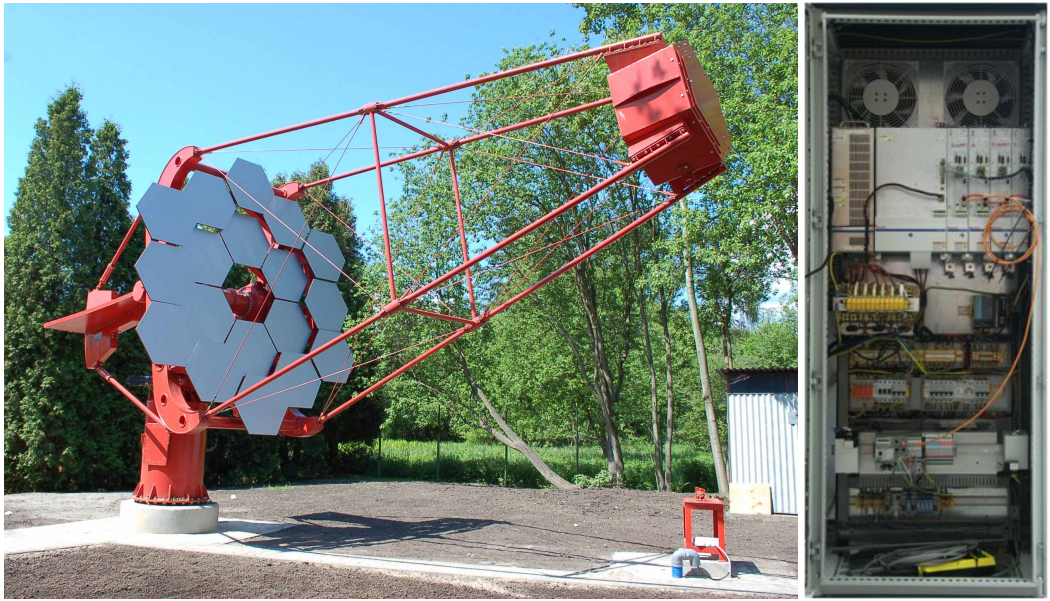

**Figure 1:** *Left:* Mechanical prototype of the SST-1M telescope at INP PAS site in Kraków. *Right:* Control cabinet for the drive and safety system.

The drive control and safety system of the SST-1M telescope is based on the solution used in the MST. This has been done in order to optimize the development effort and ease the long term maintenance of the software. The system is based on two PLCs: the main PLC dedicated to the drive system movement and a so-called Safety PLC. The latter controls and supervises all drive operations, handles the emergency procedures, and oversees the starting and stopping procedures of the camera and its safety. The main PLC driver and the Safety PLC are connected with the OPC-UA server of the Array Control Unit (ACTL). This allows for a remote control at ACTL level that will provide timestamped sequences of targets' Alt/Az positions.

The safety system includes both hardware and logic implementation to ensure human and telescope safety at the same time. Three types of safety limits (software limit, electronic and mechanical limit switches) are implemented in the SST-1M prototype to prevent movement of the telescope to the forbidden positions. For drive control each of the main telescope axis is equipped with one multi-turn external encoder. The control software is already developed and undergoes tests to ensure its full functionality. The drive control program is fully compatible with the MST program. Its logic provides inputs which are used to change the settings respectively for the MST and the SST-1M. In order to have a standardized communication protocol, all drive system related commands and their inputs and outputs have been defined. These commands – the same as those used for MSTs – will be used in the final PLC code.

## 3. FEM Analysis

Basic conclusions of the Finite Element Method (FEM) analysis [1] were that the telescope structure is quite robust and can sustain wind speeds of 200 km/h and severe seismic activities. Here we quote results of additional work that are relevant for the prototype tests described in Section 4.

To assure a proper rigidity and tension in the mast when the telescope is subject to an arbitrary external load force, the rods in the mast need to be pre-stressed to about 5100 N. The five lowest normal frequencies of the telescope frame at 45° elevation angle then read 3.46 Hz, 3.86 Hz, 9.44 Hz, 9.81 Hz, and 11.1 Hz. A relative deflection of the camera with respect to the dish during normal operation and between 45° and 95° elevation angles is 1.5 mm. For gravity, wind loads (V=36 km/h), and thermal load combined, the total deflections are not greater than 5 mm, and thus are well within 1/3 of the physical pixel size limit (8.1 mm). The optical performance degrades only by 10% with increasing the average wind speed to V=100 km/h.

## 4. Tests of the prototype structure

Several measurements and tests of the telescope prototype have been performed so far. Here we summarize the methods used and the main results.

### 4.1 Pre-stress setting in the mast rods

The method to apply the pre-tensioning of the mast rods to about 5100 N (Section 3) used Vishay tensometers (strain gauges) that were mounted on all 16 link elements. The signal from the strain gauges was registered and analyzed in live-time by the recorder. The procedure was performed at 90° elevation angle. It was followed by a series of test measurements, during which the readings were taken while the telescope was put first into a horizontal position and then back again into the zenith position. It was thus verified that no hysteresis effects occur and the stress in the rods stays within the expected range at all elevation angles. In particular, for the mast in the horizontal position, the strain in all rods is such that no compressive stresses are allowed, i.e., the tension force is always greater than about 3500 N for the rods in the bottom part of the mast to which compressive forces are applied due to the mass load of the mast.

### 4.2 Surveying measurements

The aim of the surveying measurements was to verify the parameters of the telescope structure against the CTA specifications regarding the error budget on the telescope focal length, on the relative deformation of the camera and the mirror dish, on the relative orientation of the telescope axes, the reproducibility of the mechanical deformations, and the stability of the telescope foundation. Natural modes of the telescope vibrations have been also analyzed with the use of the interferometric radar.

*Structure deformations*
Structure measurements have been done with the use of a network of 4 electronic total stations. 20 points on the telescope structure (Figure 2), placed symmetrically on both sides, have been measured at different elevation and azimuth angles. Points 1 through 4 and 21 through 24 were chosen in parts of the structure that were assumed to suffer very little deformations. Subsequently, these data points served as a reference for the determination of the static deformations of the mast structure. With the methods adopted, the mean accuracy of the measurements was ±0.5 mm. The measured deformations of the telescope were considered to be meaningful if their value was at least three times the accuracy value (±1.5 mm). The main results are as follows:

1. Relative deformations between the camera and the mirror dish in the full elevation range reach 2.5 mm. As shown in Figure 3, these deformations increase with a change in the elevation angle from 95° down to the parking position (-13°). Relative deflections between 45° and 95° elevation angles are in the range from -2.1 mm to -1.7 mm, in good agreement with a 1.5 mm value estimated through FEM calculations. Within the accuracy of the measurements, the telescope structure does not suffer any deformations with the change of the azimuth angle.

2. The reproducibility of the mechanical deformations is very good. It was checked by surveying several times the same points on the structure approached from different driving distances. The data also allowed us to estimate that the precision of the telescope control is better than 1'.

3. A relative orientation of the azimuth and the elevation axes is close to perfect — the angle between the azimuth and the elevation planes is equal to 90°04'50" and the tilt angle of the azimuth axis with respect to zenith is 0°05'55".

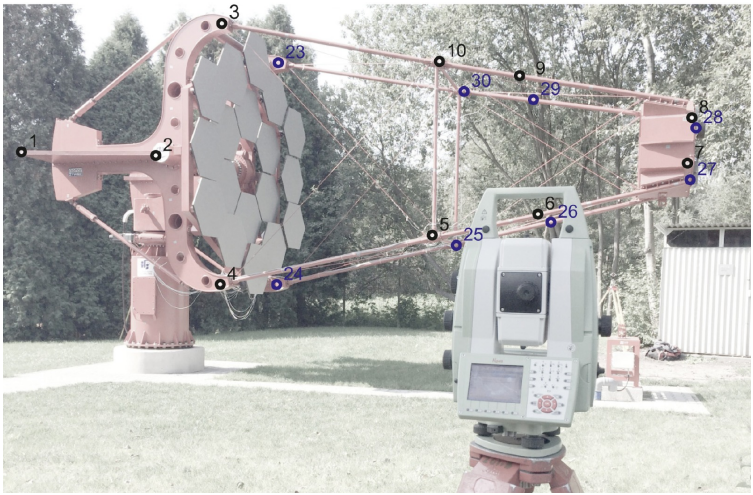

**Figure 2:** Points on the telescope frame that have been surveyed with surveying measurements – a side view.

*Natural modes of telescope vibrations*

A modal analysis of the telescope structure has been also performed with the use of the interferometric radar for 135° azimuth and 10° elevation angles (Figure 4). Natural modes identified for the telescope were at frequencies 2.76 Hz, 26.87 Hz, 32.58 Hz,..., that only slightly depend on the type and strength of the excitation. This is in good agreement with the results of the full spectral analysis performed with the use of accelerometers (see below), and also in line with our FEM calculations (Section 3). Note, however, that the surveying method did not pick up the frequencies predicted by our model above the lowest eigenfrequency and below their second peak at 26.87 Hz.

### 4.3 Full modal analysis

Natural modes of the telescope frame have been also identified through a modal experiment that used piezoelectric accelerometers and an impact hammer to excite the structure. In total, 19 modal experiments have been performed for different elevation angles (-10° (close to parking position, but not locked in docking station), -13° (telescope locked), 45°, and 90°), different points of excitation (dish support structure, mast, camera box, counterweight), and different directions of the excitation applied (X, Y, or Z) – see Figures 1, 2, 3. In addition, some experiments were done with the drives turned on (typical situation during observations), and some with the brakes on (drives off at parking position). The reference frame assumed originates at the base of the telescope tower, with the X-axis along the telescope mast, and the Z-axis in the vertical direction. The results are valid in the frequency range between 1.9 Hz and 32 Hz.

Tables 1, 2, and 3 show 5 lowest natural frequencies for the three elevation angles and 18 cases investigated when the telescope is not locked in the docking station. The lowest frequency was detected at 90° elevation angle and reads 2.61 Hz. However, typical values of the first mode frequency are close to 2.8 Hz. While the choice of the direction and the point of the excitation does not significantly change the observed frequencies, at some excitation cases certain modes were not visible. A detailed analysis also revealed that natural frequencies of the telescope with the drives turned on are about 0.2 Hz lower than the frequencies measured with the brakes on. Note that these

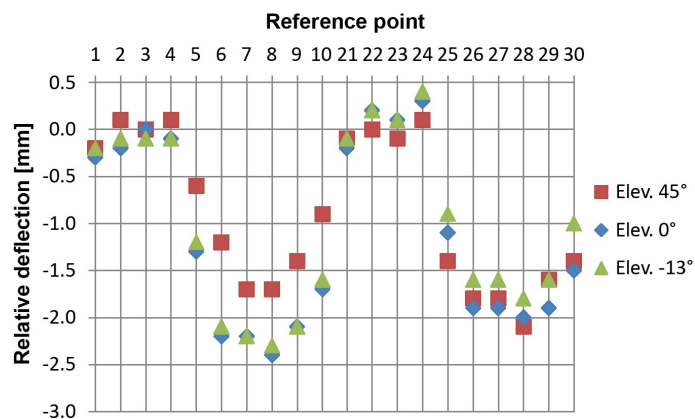

**Figure 3:** Deflections in the telescope structure measured at the survey points (see Fig. 2) at different elevation angles relative to the deflections at 95° in the elevation angle. Points 1-4 and 21-24, in which the telescope structure exhibits negligible deformations, formed a reference for the spatial transformations that led to these results.

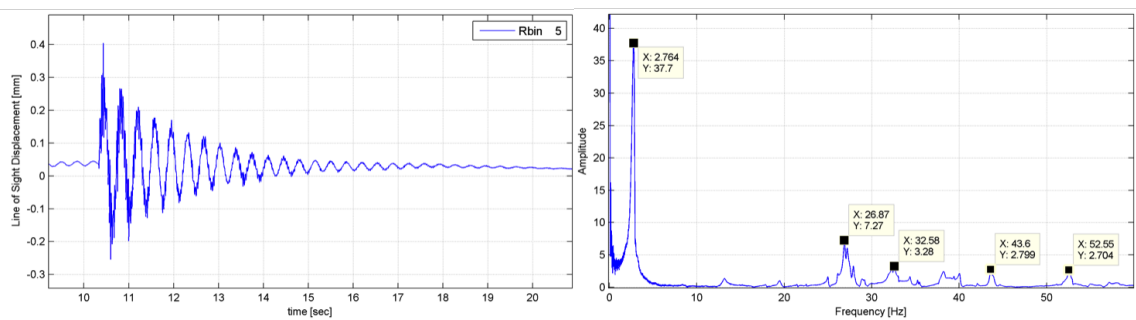

**Figure 4:** The line of sight displacement (left) and the corresponding amplitude spectrum (right) for one of the excitation analyzed for the telescope at 135° azimuth angle and 10° elevation angle with the use of the interferometric radar.

results are in good agreement with the measurement taken with geodesic methods. The measured values are somewhat lower than the ones obtained in our FEM calculations (Section 3) due to simplifications applied in the finite-element model. The last, 19th case analyzed was with the telescope locked in the docking station and the excitation at the camera box in the Z-direction. As expected, the natural frequencies increase in such a case. Five lowest frequencies of the structure read: 4.03 Hz, 5.39 Hz, 6.68 Hz, 7.85 Hz, and 9.23 Hz.

| Exc. Point | dish supp. | mast | mast | camera | camera | camera |
|---|---|---|---|---|---|---|
| Exc. Direction | Z | Z | Y | Z | X | Z |
| Elevation | | | -10° | | | |
| Drive on/off | off | off | off | off | off | on |
| Exp.# | P1 | P2 | P3 | P4 | P5 | P18 |
| Natural frequencies [Hz] | | | | | | |
| 1 | 2,98 | 2,86 | | 2,86 | 3,00 | 2,74 |
| 2 | 3,47 | 3,42 | 3,46 | 3,39 | 3,46 | 3,35 |
| 3 | | | 3,72 | | | |
| 4 | | | | | | 5,03 |
| 5 | 6,61 | 6,59 | 6,59 | 6,63 | 6,61 | 6,55 |

**Table 1:** Five lowest natural frequencies of the telescope frame at $-10°$ elevation angle for modal experiments that differ in the excitation point and direction. In cases P1-P5 the drives were turned off.

| Exc. Point | dish supp. | dish supp. | counterweight | counterweight | mast | mast |
|---|---|---|---|---|---|---|
| Exc. Direction | Y | Z | Y | Z | Y | Z |
| Elevation | | | 45° | | | |
| Drive on/off | on | on | on | on | on | on |
| Exp.# | P6 | P7 | P8 | P9 | P10 | P11 |
| Natural frequencies [Hz] | | | | | | |
| 1 | | 2,97 | 3,16 | 2,92 | | 3,06 |
| 2 | 3,46 | | 3,45 | | 3,48 | |
| 3 | 8,59 | 8,59 | 8,54 | 8,56 | 8,57 | 8,58 |
| 4 | 8,94 | 9,06 | 8,88 | 9,07 | 9,09 | 9,13 |
| 5 | 9,83 | 9,91 | 9,10 | 9,91 | 10,04 | |

**Table 2:** Natural frequencies of the telescope frame at 45° elevation angle. Compare Table 1.

| Exc. Point | mast | mast | dish supp. | dish supp. | counterweight | counterweight |
|---|---|---|---|---|---|---|
| Exc. Direction | Z | Y | Y | Z | Y | Z |
| Elevation | | | 90° | | | |
| Drive on/off | on | on | on | on | on | on |
| Exp.# | P12 | P13 | P14 | P15 | P16 | P17 |
| Natural frequencies [Hz] | | | | | | |
| 1 | | | | 2,87 | | 2,61 |
| 2 | | 3,25 | 3,46 | 3,47 | 3,46 | 3,47 |
| 3 | | 3,82 | | | | |
| 4 | | | | 6,58 | 6,68 | 6,79 |
| 5 | 7,18 | 7,09 | 7,07 | 7,03 | 7,06 | 7,08 |

**Table 3:** Natural frequencies of the telescope frame at 90° elevation angle. Compare Table 1.

## 4.4 Drive and safety system tests

The aim of the drive system tests, that are still ongoing, is to verify a proper functionality of the control software and its implementation on hardware devices. In order to verify the positioning accuracy of the telescope and to check for the existence of vibrations of the structure during slewing, a step response of the telescope has been measured for different position sets and driving distances. An example result for a target in the observation range and default settings in the control software is presented in Figure 5. The position was taken from the external encoders readings recorded every 80 ms. Time resolution thus allows a detection of any oscillation modes at the natural frequencies of the telescope structure. One can see that a required position is reached with the required accuracy of less than 0.1°. The telescope motion is smooth and synchronized and no vibrations of the structure are detected. There is also no overshoot at approaching a target position. This is due to the controlled breaking implemented in the control algorithm.

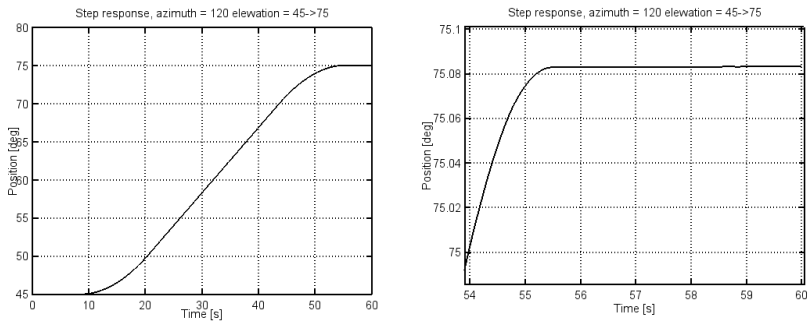

**Figure 5:** Step response of the telescope slewing from 45° to 75° in elevation and at constant azimuth angle of 120°. The right panel shows the zoom-in of the late stage of motion.

Other tests performed also show that the drive torques during tracking (with backlash elimination) and pointing well agree with the designed values. In emergency, the telescope can be safely stopped in less than 1 s. On the occasion of software development, tests, and modernization works the telescope has been operating at least 3 times a week, from 15 min to 3 h per day, usually performing fast movements with high acceleration. The drive system has remained reliable up to date. Additional regular automated tests with positioning and tracking motions are planned to be started in the near future, for which a remote control board operation has been already implemented. These tests will prove a long-term performance of the drive system.

## 5. Summary

The SST-1M prototype mechanical structure has been successfully built at a test site at INP PAS in Kraków. The tests performed so far prove that SST-1M is viable solution for the small-size telescope of CTA.

## Acknowledgments

We gratefully acknowledge support from the agencies and organizations listed under Funding Agencies at this website: http://www.cta-observatory.org. In particular we are grateful for support from the Polish NCN grant No. DEC-2011/01/M/ST9/01891, the Polish MNiSW grant No. 498/1/FNiTP/FNiTP/2010, the University of Geneva, and the Swiss National Foundation.